# Extracting Knowledge From Massive Astronomical Data Sets

Massimo Brescia[a], Stefano Cavuoti[b], S. G. Djorgovski[c,d], Ciro Donalek[c],
Giuseppe Longo[b,e], Maurizio Paolillo[b]

[a]*INAF – Osservatorio Astronomico di Capodimonte, Via Moiariello 16, 80131 Napoli, Italy*
[b]*Dipartimento di Fisica, Università degli Studi Federico II, Via Cintia 26, 80125 Napoli, Italy*
[c]*California Institute of Technology, 1216 East California bvd., Pasadena, CA 91125, USA*
[d]*Distinguished Visiting Professor, King Abdulaziz Univ., Jeddah 21589, Saudi Arabia*
[e]*Visiting Associate, California Institute of Technology, Pasadena, CA 91125, USA*

**ABSTRACT:**

The exponential growth of astronomical data collected by both ground based and space borne instruments has fostered the growth of Astroinformatics: a new discipline laying at the intersection between astronomy, applied computer science, and information and computation (ICT) technologies. At the very heart of Astroinformatics is a complex set of methodologies usually called Data Mining (DM) or Knowledge Discovery in Data Bases (KDD). In the astronomical domain, DM/KDD are still in a very early usage stage, even though new methods and tools are being continuously deployed in order to cope with the Massive Data Sets (MDS) that can only grow in the future. In this paper, we briefly outline some general problems encountered when applying DM/KDD methods to astrophysical problems, and describe the DAME (DAta Mining & Exploration) web application. While specifically tailored to work on MDS, DAME can be effectively applied also to smaller data sets. As an illustration, we describe two application of DAME to two different problems: the identification of candidate globular clusters in external galaxies, and the classification of active galactic nuclei (AGN). We believe that tools and services of this nature will become increasingly necessary for the data-intensive astronomy (and indeed all sciences) in the 21$^{st}$ century.

## 1. Introduction

Over the last decade or two, due to the evolution of instruments and detectors, astronomy has become an immensely data rich science, thus triggering the birth of Astroinformatics: a new discipline placed at the crossroad between traditional astronomy, applied mathematics, computer science and ICT technologies. Among the other things, Astroinformatics aims at providing the astronomical community with a new generation of accurate and reliable methods and tools needed to reduce, analyse and understand massive and complex data sets and data flows which go far beyond the reach of traditionally used methods.

A good example of the challenges that have to be addressed by the astronomical community is the Large Synoptic Survey Telescope (LSST, [1]) which should become operational within this decade, and which will produce a data flow of about 20 – 30 TB per observing night, or many PB/year. LSST raw data will therefore need to be calibrated, analyzed and processed in real time and, speed, accuracy and reliability become a must. This need is made even more urgent by the fact that fast digital surveys allow multiple passes over large patches of the sky thus leading to the discovery of large numbers of photometric and astrometric transients. LSST, for instance, is expected to find $10^5 – 10^6$ variable or transient objects per observing night. Most of these objects will need to be immediately recognized and classified in order to implement proper follow-up observations [38,39].

It is worth emphasizing that most astronomical data archives are currently being federated and made interoperable by the various Virtual Observatory projects; see International Virtual Observatory Alliance (IVOA) [2]. It is apparent that nowadays, and even more in the future, the most interesting problems will call for the use of complex, multi-wavelength, multi-epoch data collected with heterogeneous instruments. However, the extraction of useful and relevant knowledge from such datasets is still a highly non-trivial task which requires a new generation of software tools: automatic, scalable and highly reliable. This has been recently recognized by the implementation of a specific Interest Group on Knowledge Discovery in Databases within the IVOA [2]. In what follows we shall focus on some recent developments in the field of astronomical Data Mining (hereafter DM) or Knowledge Discovery in Databases (KDD). Some early reviews of the topic include, e.g., [3, 4, 31, 32, 33, 34, 35, 36, 37].

## 2. Astronomical KDD

In the broadest sense, KDD/DM regards the discovery of "models" for data. There are, however, many different methods which can be used to discover these underlying models: statistical pattern recognition, machine learning, summarization, etc., and an extensive review of all these models would take us far beyond the purposes of this paper. In what follows we shall therefore sumarrise only the main methodological aspects. For a detailed analysis of the methods used to derive models for data, the general reader can refer to [5,6]. Machine learning (ML), which is sometimes considered to be a branch of artificial intelligence (AI), is a scientific discipline concerned with the design and development of algorithms that allow computers to evolve behaviors based on empirical data. A "learner" can take advantage of examples (data) to capture characteristics of interest of their unknown underlying probability distribution.

These data form the so called Knowledge Base (KB): a sufficiently large set of examples to be used for training of the ML implementation, and to test its performance. The difficulty lies in the fact that often, if not always, the set of all possible behaviors given all possible inputs is too large to be covered by the KB. Hence the learner must possess some generalization capabilities in order to be able to produce useful output when presented new instances.

From a completely general point of view, regardless the specific method implemented, DM is a rather complex process. In most cases the optimal results can be found only on a trial and error base by comparing the outputs of different methods or of different implementations of the same method. This implies that in order to solve a specific problem a lengthy fine-tuning phase is often required. Such complexity is among the reasons for a slow uptake of these methods by the community of potential users which still fail to adopt them. In order to be effective, a DM application requires a good understanding of the mathematics underlying the methods, of the computing infrastructure, and of the complex workflows which need to be implemented. So far, most domain experts in the scientific community are simply not willing to make the effort needed to understand the fine details of the process, and prefer to recur to traditional approaches which are far less powerful, but which may be more user-friendly. This situation is unsustainable as the ever larger MDS become available, and there will be no viable alternatives to DM methods for their exploration.

Here we shall adopt a definition of DM based on the fact that in most real-life applications (hereafter "use cases") several different methods (hereafter called "functionalities") need to be combined to produce a reliable and robust result. A possible taxonomy of DM functionalities is:
- dimensionality reduction
- classification
- regression
- prediction
- clustering
- filtering

Each functionality can be implemented with a variety of models (e.g., Neural Networks, Support Vector Machines, Bayesian Networks, etc.), and use cases are therefore defined by a proper association of the functionality+model. This taxonomy is at the foundation of the DAME (Data Mining & Exploration) web application described below.

Put simply, classification is a procedure in which individual items are placed into groups based on some quantitative information (referred to as features) using the knowledge contained in a training set of previously labeled items. A classifier is therefore a system that performs a mapping from a feature space **X** to a set of labels **Y**. Classification may be either "crispy" or "probabilistic". In the first case, given an input pattern **x** the classifier returns its computed label **y**. In the second case, given an input pattern **x** the classifier returns a vector **y** which contains the probability of $y_i$ to be the "right" label for **x**. Both types of classification can be applied regardless of the number of classes considered. Typical astrophysical problems which have been addressed with this functionality are the so called star-galaxy separation (which would be better called resolved-unresolved objects separation), morphological classification of galaxies, automatic classification of stellar spectra, etc.

Regression is defined as the supervised search for a mapping from a domain in $R^n$ to a domain in $R^m$, where m < n (where n is the number of input features and m the number of output variables). One can distinguish between two different types of regression: (1) data table statistical correlation, in which the user tries to find a mapping without any prior assumption on the functional form of the data distribution; or (2) function fitting: in which the user tries to validate the hypothesis, suggested by some theoretical framework, that the data distribution follows a well defined, and known, function. The most common astrophysical example of a regression problem is the evaluation of photometric redshifts of galaxies.

Finally, clustering techniques apply when there is no class to be predicted, but rather when the instances need to be divided into "natural" (e.g., statistically justified) groups. Since no extensive KB is required, this methods are also referred to as unsupervised models [6].

In general there are different ways in which the results of clustering can be expressed: for instance, the identified groups can be exclusive or overlapping. They may be also probabilistic, whereby an instance belongs to each group with a certain probability. Some clustering algorithms produce a hierarchical structure of clusters, so that at the top level the instance space divides into just a few clusters, each of which divides into its own sub-clusters at the next level, and so on. Clustering is sometimes followed by a stage where a decision tree or "set of rules" is inferred in order to allocate each instance to the cluster to which it belongs. The choice between these different models is dictated by the nature of the specific problem to be tackled.

In spite of the enormous potential (for example, a possible identification of the previously unknown types of objects in the parameter space), the application of clustering methods to astrophysical MDS is still in a very early stage, although in some cases [e.g., 7] they are embedded into complex DM workflows.

Before proceeding further, we address briefly how these DM functionalities and methods can be applied to different datasets.

## 3. Computational Requirements

Without entering into too much detail, we note that DM on MDS poses two important challenges for the computational infrastructure: asynchronous access and scalability.

Most available web-based DM services run synchronously, i.e., they execute jobs during a single HTTP transaction. This may be useful and simple, but it does not scale well when it is applied to long-run tasks. With synchronous operations, all the entities in the chain of command (client, workflow engine, broker, processing services) must remain up for the duration of the activity: if any component stops, the context of the activity is lost.

Regarding scalability, whenever there is a large quantity of data, there are three approaches to making learning feasible. The first one is trivial, consisting of applying the training scheme to a decimated data set. Obviously, in this case, the information may be easily lost and there is no guarantee that this loss is negligible in terms of correlation discovery. This approach, however, may turn very useful in the lengthy optimization procedure that is required by many ML methods (such as Neural Networks or Genetic Algorithms).

The second method relies in splitting the problem in smaller parts (parallelization) sending them to different CPUs and finally combine the results together. However, implementation of parallelized versions of learning algorithms is not always easy [8], and this approach should be followed only when the learning rule, such as in the case of Genetic Algorithms [9], or Support Vector Machines [10], is intrinsically parallel. However, even after parallelization, the asymptotic time complexity of the algorithms cannot be improved.

A third and more challenging way to enable a learning paradigm to deal with MDS is to develop new algorithms of lower computational complexity, but in many cases this is simply not feasible [e.g., 11].

In some situations, background knowledge can make it possible to reduce the amount of data that needs to be processed by adopting a particular learning rule, since in many cases most of the measured attributes may turn out to be irrelevant or redundant when background knowledge is taken into account [5,6]. In many exploration cases, however, such background knowledge simply does not exists, or it may introduce biases in the discovery process.

## 4. DAME Web Application

The DAME web application [13] is a joint effort between the Astroinformatics groups at University Federico II, the Italian National Institute of Astrophysics, and the California Institute of Technology.

DAME aims at solving in a practical way some of the DM problems, by offering a completely transparent architecture, a user-friendly interface, and the possibility to seamlessy access a distributed computing infrastructure. DAME adopts VO standards in order to ensure the long-term interoperability of data; however, at the moment, it is not yet fully VO compliant. This is partly due to the fact that new standards need to be defined for data analysis, DM methods and algorithms development. In practice, this implies a definition of standards in terms of ontology and a well-defined taxonomy of functionalities to be applied in the astrophysical use cases.

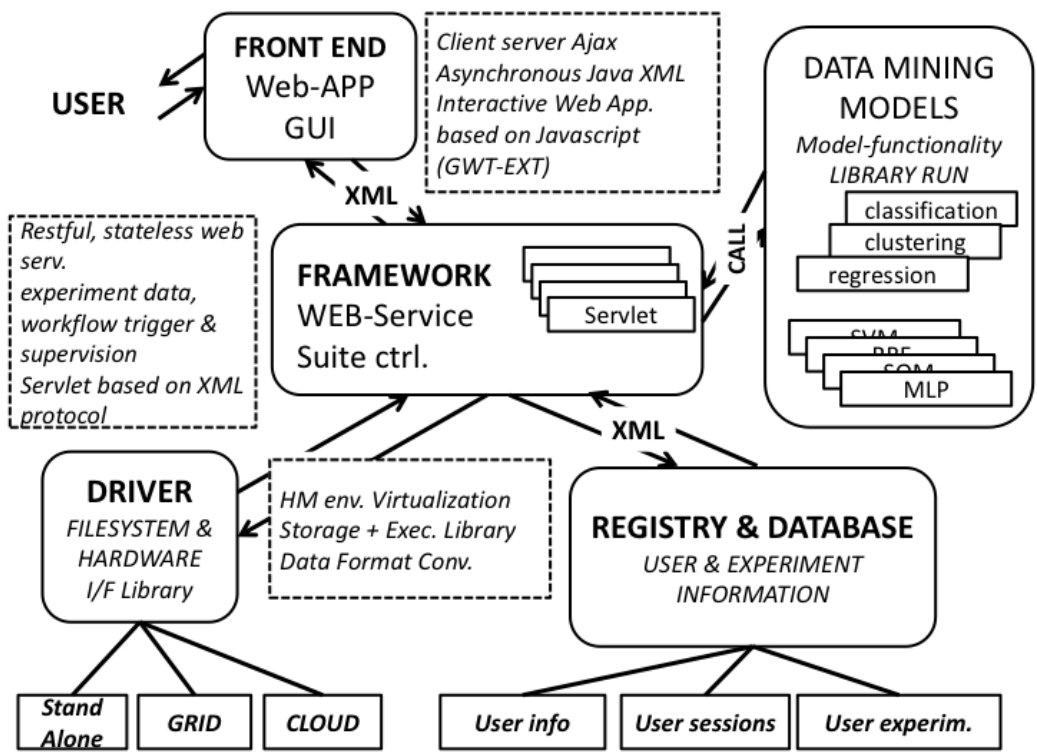

**Figure 1.** The DAME functional structure

In order to effectively deal with MDS, DAME offers asynchronous access to the infrastructure tools, thus allowing the running of activity jobs and processes outside the scope of any particular web application operation and without depending on the user connection status. The user, via a simple web browser, can access the application resources and can keep track of his jobs by recovering related information (partial/complete results) without having the need to maintain open the communication socket. Furthermore, DAME has been designed to run both on a server and on a distributed computing infrastructure (Grid or Cloud).

The Front End component includes the GUI (Graphical User Interface). It contains dynamical web pages which are used by the end users to interact with the applications, models, and facilities to launch scientific experiments. The interface includes an authentication procedure which redirects the user to a personal session environment, where he can find uploaded data, check the experiment status and driven procedures, configure and execute new scientific experiments. This mechanism was also required for Grid access security reasons.

A detailed technical description of the other components can be found in [13]. Here we just want to explicitly notice that in the currently available DAME release (beta version, mid-2011), the

DMM implements MultiLayer Perceptron trained by three different learning rules (Back Propagation, Genetic Algorithm, Quasi Newton) and Support Vector Machines (SVM) as supervised models, Self Organizing Feature Maps (SOFM) and K-Means as unsupervised models (one more unsupervised model, the Principal Probabilistic Surfaces, PPS, is almost ready to be integrated in the web application).

However, given the large variety of DM models and taking into consideration the fact that specific problems may require the implementation of specific models that are not available in the main release, DAME includes also a Java-based plugin wizard for custom experiment setup (hereinafter called DMPlugin). In other words, the DMPlugin interface makes the web application *pluggable*, in the sense that it allows any user to upload into the DAME suite his own data mining code and run it on the computing infrastructure.

## 5. Two Use Case Examples

Here we describe the results of two problems tackled with DAME, one with a small data set, and the second on a MDS, in order to illustrate its scientific uses.

### *5.1 Selection of candidate globular clusters*

As we shall demonstrate, the use of properly tuned DM algorithms can yield very complete datasets with low contamination even with single band photometry, thus minimizing the observing time requirements and allowing to extend such studies to larger areas and to the outskirts of nearby galaxies. The detailed experiments are described in [16]. Here we shall just summarize the main aspects of this use case.

For the experiments we used a catalogue extracted (as described in [17]) from a 100 arcmin$^2$ mosaic centered on the galaxy NGC 1399, obtained with the *HST Advanced Camera for Surveys* (Fig. 2), in the F606W filter. The final catalogue contains 12,915 sources and reached a 7-σ depth of $m_V = 27.5$ mag, i.e., 4 mag below the GC luminosity function turnover, thus allowing to sample the entire GC population. The NGC 1399 region covered by the HST mosaic had no complete color coverage, and the auxiliary information needed to implement the KB on which to train the methods was obtained using archival HST (*g-z*) observations [18] covering the very central region of the galaxy (less than 10% of the sample) and (*C-T$_1$*) ground based photometry from [19]. The total number of objects in the KB was 2100.

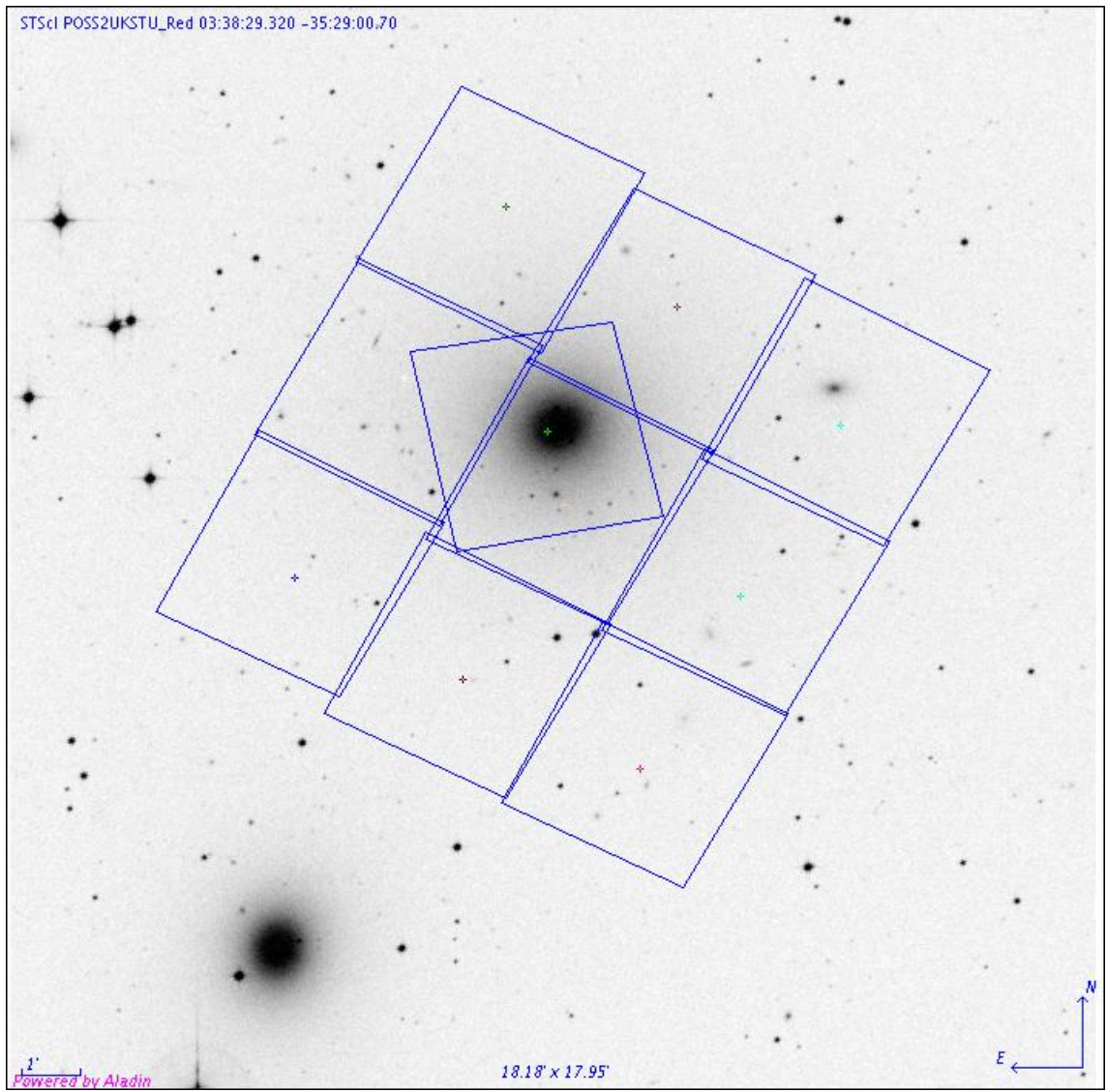

**Figure 2.** The field of view covered by the 3x3 HST/ACS mosaic in the F606W band. The central field, with a different orientation, shows the region covered by previous archival ASC observations in g and z bands.

In order to find the best model, five supervised methods were tested: three different implementation of the Multi Layer Perceptron (MLP), trained by Back Propagation (MLPBP, [5]), with Genetic Algorithm pruning of the weights (MLPGA, [9]) and with Quasi Newton learning rule (MLPQNA, [40]); Support Vector Machines (SVM, [28]); and Genetic Algorithms [41]. The details of the specific implementations can be found in [5, 12,16]. As it can be seen in Table 1, the most performing is the MLPQNA which, using all available features led to a 98.3% success rate, and using only the optical parameters to a success rate of 94.7%. This small difference (only 3%) in performance is quite surprising since it implies that the structural parameters convey only a limited amount of information. Moreover, the pruning in the mixed cases (obtained by excluding some structural and optical features) revealed an homogeneous behavior in all models, in terms of quantity of correlated information introduced by individual features in the patterns. More specifically: five optical features (the isophotal magnitude, the three aperture photometry and the FWHM of the PSF) were recognized as the most relevant by all models.

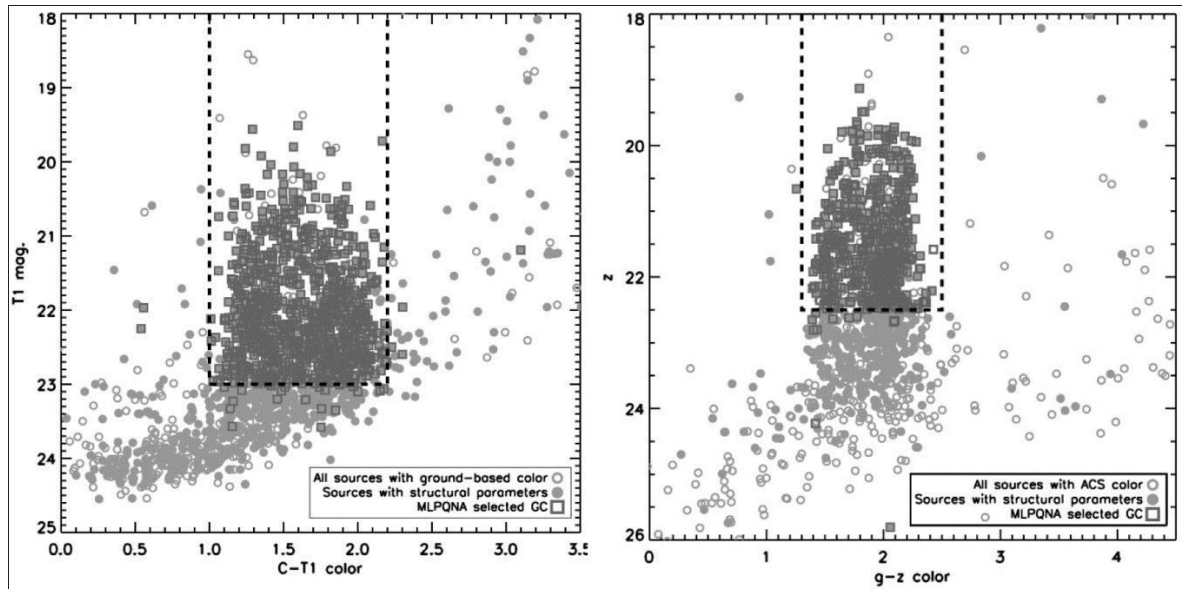

**Figure 3.** Color-magnitude diagrams using (C−T1) ground-based (left panel) and (g−z) HST photometry (right panel). Ground-based photometry covers the whole FOV of our ACS mosaic, while HST colors are limited to the central ACS field (<200″×200″, Figure 2). Open grey dots represent all sources in color catalogs while solid ones refer to the subsample with both color and structural parameters. Squares mark the MLP selected sample. The MLP sample reproduces the properties of the color-selected GC population (dashed line) with little contaminants and/or outliers.

| Experiment | Missing features | MLPQNA | GAME | SVM | MLPBP | MLPGA |
|---|---|---|---|---|---|---|
| All features |  | 98.3 | 82.2 | 90.5 | 59.8 | 66.2 |
| No feature 11 | 11 | 97.9 | 81.9 | 90.5 | 59.0 | 62.4 |
| Only optical features | 8, 9, 10,11 | 93.9 | 86.4 | 90.9 | 70.3 | 76.2 |
| Mixed features | 5,8,9,10,11 | 94.7 | 86.7 | 89.1 | 68.6 | 71.5 |

**Table 1.** Performance of the five supervised learning models tried on the GC data set.

## 5.2 Classification of AGN

Galaxies containing Active Galactic Nuclei (AGN) do not fit easily inside any morphological classification, and their identification requires usually two steps: the selection of suitable candidates in a color parameter space, and their subsequent confirmation via spectroscopy. Spectroscopic confirmation takes place by comparing the ratios (measured in terms of their equivalent width) of specific emission lines with theoretical or semi-empirical curves such as those introduced by Heckman [20], Kauffman [21], and Kewley [22].

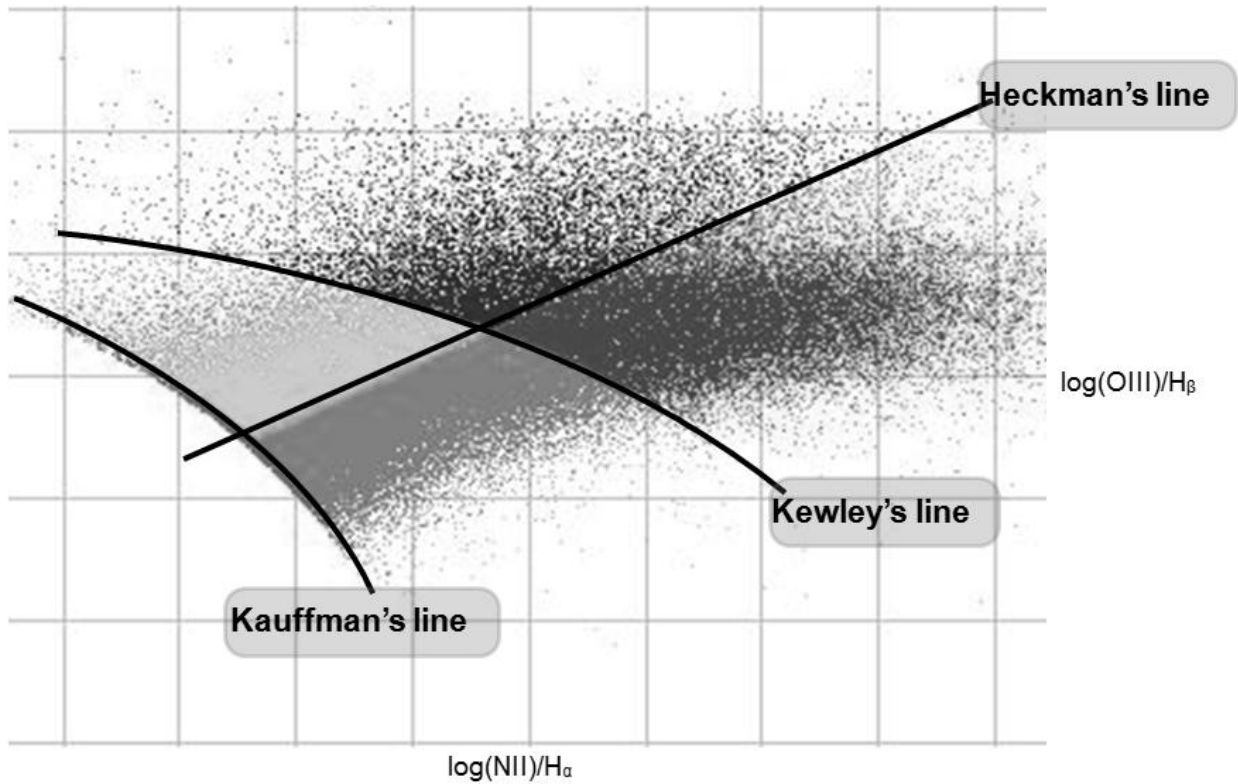

**Figure 4.** The KB used for the AGN experiment.

The selection of AGN based on photometric parameters only would allow: *(i)* saving a precious observing time; *(ii)* building larger and more complete catalogues of AGN. As it will be shown, Data Mining methods seem to offer an alternative and effective way to select AGN-hosting galaxies using just photometric parameters and a spectroscopic KB.

For our preliminary experiments (for details see [23,24]) we used data extracted from the 4-th data release of the Sloan Digital Sky Survey [24] (*PhotoObjAll* table), and a KB formed by the union of objects: residing in different regions of the BPT plot shown in Fig. 4 [25]; the AGN contained in [26] and those from the emission line catalogue in [27].

The SDSS parameters which we used as input features were: the Petrosian radii (at 50%) in the five SDSS bands, the concentration index in the *r* band, the *fibermag_r*; the dereddened colors and the dereddend magnitude in the *r* band, plus the photometric redshifts obtained in [27].

For the experiments we used two types of models: MLPBP [5], and Support Vector Machines (LIBSVM library [28]). The latter implementation provides 2 methods for classification and two methods for regression. SVM imply the search for a minimum in the space defined by at least two hyperparameters. The basic idea behind SVM is that they look for a set of hyperplanes partitioning the parameter space in regions (i.e., classes) where the separation between different classes (as defined by the objects in the KB) is maximum. In the case of a two classes classification problem it is necessary two determine two parameters (called hyperparameters and, in the case of the C-SVM model, conventionally labeled as *C* and $\gamma$).

Following the approach proposed in [29] we run different jobs on a grid whose knots were spaced by a factor 4 on both parameters ($\gamma = 2^{-15}, 2^{-13}...2^3, C = 2^{-5}, 2^{-3}, ...2^{15}$) thus leading to a total of 110 processes for each experiment. Cross-validation of results and "folding" (5 subsets) of the dataset were used for all experiments. The goal of the experiments was conservative since we were not interested in completeness, but rather in minimizing the fraction of false positives. Three experiments were performed:

- Experiment 1: classification AGN/non-AGN.
- Experiment 2: Type 1 AGN/Type 2 AGN as defined in [26]: if FWHM(H$\alpha$)> 1.5 FWHM([OIII]l5007) or (FWHM(H$\alpha$) > 1200 km s$^{-1}$ and FWHM([OIII]l5007) < 800 Km s$^{-1}$), then the AGN is considered of Type 1 otherwise it is of type 2.
- Experiment 3: Seyfert galaxies / liners.

| experiment | features | KB | algorithm | $e_{tot}$ | C (MLP only) |
|---|---|---|---|---|---|
| AGN vs. non-AGN | SDSS photom. + photo-z | BPT plot + Kewley's line | SVM | 74 % | AGN ca. 55% non-AGN ca. 87% |
| | | | MLP | 76 % | |
| Type 1 AGN vs. Type 2 | SDSS photom. + photo-z | Sorrentino et al. + Kewley line | SVM | $e_{type1} = 82\%$ $e_{type2} = 86\%$ | Type 1 = 99% Type 2 = 100% |
| | | | MLP | $e_{type1} = 99\%$ $e_{type2} = 98\%$ | |
| Seyferts vs. Liners | SDSS photom. + photo-z | BPT + Heckman & Kewley lines | SVM | 78 % | Seyfert ca. 53% Liners ca. 92 % |
| | | | MLP | 80 % | |

**Table 2.** Summary of the performances of SVM and MLP on the AGN classification task. Column1: experiment; (2) features utilized for the experiments; (3) type of KB; (4) method; (6) total efficiency for the two algorithm; (7) completeness for the two class derived with the MLP.

## 6. Conclusions

In practice, the most interesting DM problems do not come from computer scientists, or from machine learning experts, or from the data themselves, but from people and communities who tackle specific problems. Usually, these domain experts find it difficult or too time-consuming to master the DM methodology that is intrinsically complex. Applications and services like DAME can make the use of advanced DM methods and workflows as user friendly as possible, in order to empower scientists who are not ML experts to apply these techniques to their specific problems.

Quoting from [4], "*Astronomers are not cutting-edge experts in data mining algorithms any more than they are in statistics, databases, hardware, software, etc., but they will need to know enough to usefully apply such approaches to the science problem they wish to address. It is likely that such progress will be made via collaboration with people who are experts in these areas, particularly within large projects, that will employ specialists and have working groups dedicated to data mining.*" Many issues, however, remain still to be solved before effective and seamless DM on huge and distributed data sets becomes a standard research practice.

First of all, the available computing infrastructure should be designed to optimize (or at least enable) a new DM approach to research. Second, in order to allow mining of multiple or distributed data sets (e.g., after crossmatching of data in different archives) these data sets need to be made interoperable at an higher level than what is now generally available within the standards provided by the VO. It is possible that the VOSpace protocol [39] will at least in part solve this problem. Since it is unlikely that most researchers will either require or trust the advanced DM tools, it is useful to: (i) provide the use of standard, well-understood and trustesd methods for an exploratory work; (ii) allow for a possibility that the user can run their own or trusted code on their own data. Both of these are at least partly solved within the DAME environment.

For MDS, scalability of DM algorithms is still an open issue. In the case of DAME, for instance, the implementation of parallel codes on GPUs via CUDA programming rules might help to further improve the scalability of some models, while the interoperability with other data mining infrastructures, such as KNIME [29], can be used to achieve a better interoperability between data representation and mining.

In conclusion, we believe that a broader uptake and application of DM/KDD tools for the data-rich science in the 21$^{st}$ century is both necessary and inevitable, driven by an ever growing need for an effective knowledge extraction from massive and complex data sets. DAME is an example of a scientific research tool, made publicly available as a web facility, that may facilitate the adoption of these modern methods in the astronomical community and beyond.

## Acknowledgments

The DAME web application has been funded in part by the Italian Ministry of Foreign Affairs through a great relevance bilateral projects between Italy and USA, and by the Italian Ministry of Research and University through the PON 1575 S.Co.P.E. SGD and CD acknowledge a partial support through the NASA grant 08-AISR08-0085, the NSF grants AST-0834235 and AST-0909182, and the Fishbein family foundation. We thank numerous collaborators for many interesting discussions on these and related issues over the years.